# Prompt-to-prescription: towards generative design of diffraction-limited refractive optics

*Roy Maman[1]\*, David Ohana[1], Jacob Engelberg[1,2] and Uriel Levy[1]*

1- Institute of Applied Physics, The Faculty of Science, The Hebrew University of Jerusalem, Jerusalem 91904, Israel
\*E-mail: Roy.maman@mail.huji.ac.il.
2-Department of Electro-Optics and Applied Physics, Jerusalem College of Technology, 9116001, Jerusalem, Israel.

## Abstract

The design of high-performance optical systems remains a specialized domain gated by the limited availability of expert engineers, creating a bottleneck that stalls innovation despite the growing demand for imaging hardware. While deep learning has improved parameter optimization, it has yet to address the fundamental challenge of conceptualizing valid optical architectures from functional requirements. Here, we present an end-to-end generative framework that couples the semantic reasoning of Large Language Models (LLMs) with a differentiable ray-tracing engine to democratize the synthesis of diffraction-limited optical prescriptions. By treating optical design as a semantic-to-physical translation task, the system autonomously interprets prompts ranging from high-level end-user requests to rigorous technical specifications. We demonstrate the framework's versatility across three distinct regimes: (1) finite-conjugate industrial metrology systems, where the model autonomously enforces application-specific constraints such as telecentricity to achieve diffraction-limited performance; (2) a suite of infrared objectives (NIR, SWIR, and LWIR), demonstrating the framework's ability to synthesize valid topologies and optical prescriptions for non-visible spectral bands, and (3) complex aspheric mobile lenses, where the system successfully navigates the high-dimensional optimization landscape to produce high-resolution designs suitable for modern sensors. Validated against industry-standard simulation tools, these results establish a new paradigm for automated optical engineering, bridging the gap between semantic intent and physical realization.

## Introduction

Modern optical systems are increasingly required to meet demanding and often competing constraints on size, performance, manufacturability, and integration. Applications ranging from compact imaging modules and augmented reality to silicon photonics, atomic sensors, and spaceborne instrumentation routinely require multi-element optical designs operating close to physical limits. Traditionally, designing such systems has been a highly iterative process, relying heavily on expert intuition, manual specification, and local numerical optimization within closed-source commercial software.

Recent advances in differentiable ray tracing[1–3] and gradient-based optical optimization have enabled direct, physics-aware optimization of complex optical systems under realistic constraints. In parallel, large language models (LLMs) have demonstrated a remarkable ability to encode and reproduce structured technical knowledge from natural language descriptions[4–

[7]. This technological convergence has recently motivated a shift in the industrial landscape, with emerging commercial tools seeking to integrate learning-based modules into the optical engineering workflow. However, as these platforms move toward proprietary, 'black box' implementations focused primarily on end-user productivity, there remains a critical need for a transparent, scientifically grounded framework, that establishes the fundamental link between linguistic intent and physical realization.

Prior work has explored neural networks for surrogate modelling[8–11] and inverse design of individual optical components[12–19]. However, most existing approaches either operate at the level of isolated elements or remain tightly coupled to specific parameterizations and optimization pipelines[20–35]. As a result, they do not address the central challenge now being validated by industry: the direct synthesis of complete, physically valid optical systems from high-level design intent expressed in natural language.

In this work, we introduce an end-to-end optical design framework that converts a natural-language prompt directly into a complete, optimizable optical prescription, compatible with standard ray-tracing software. The key insight is to treat optical design as a constrained optimization problem that becomes tractable once a physically meaningful initialization is available. An LLM is used to generate this initialization, encoding common optical design heuristics, architectural choices, and feasibility constraints. A differentiable ray-tracing engine performs subsequent physics-based optimization. Together, these components bridge the gap between human intent and numerical optimization, enabling automated optical design from high-level textual descriptions.

This work establishes a new paradigm for optical system synthesis, in which natural-language specifications serve as a first-class interface to physically grounded optical design. Beyond accelerating expert workflows, the proposed approach lowers the barrier to entry for non-specialists and opens new possibilities for automated co-design across optical, photonic, and electronic domains.

## Results

### System Architecture and End-to-End Pipeline

The architectural foundation of our design framework rests on a strategic choice between two primary AI paradigms: Supervised Fine-tuning (SFT) and Retrieval-Augmented Generation (RAG). Fine-tuning involves retraining a model's internal parameters on a specialized dataset, essentially 'teaching' the AI optical design through repetition. While this can adapt a model to technical nomenclature, it often leads to 'catastrophic forgetting', where the model loses its broader physical reasoning, and a tendency to 'hallucinate' non-physical designs when faced with novel constraints[36–38]. Consequently, we adopt a Reasoning-First RAG approach powered by Claude Sonnet 4.5 (Anthropic). Instead of modifying the model's weights, we utilize the Large Language Model (LLM) as an intelligent 'semantic controller' that navigates a curated library of ~1,700 validated optical designs. By retrieving existing, physically sound examples to guide the model's synthesis, we ensure that the generated initialization is grounded in proven optical archetypes[39]. This hybrid strategy combines the creative flexibility of generative AI with the deterministic reliability of classical engineering databases[40], providing a robust starting point for high-precision optimization.

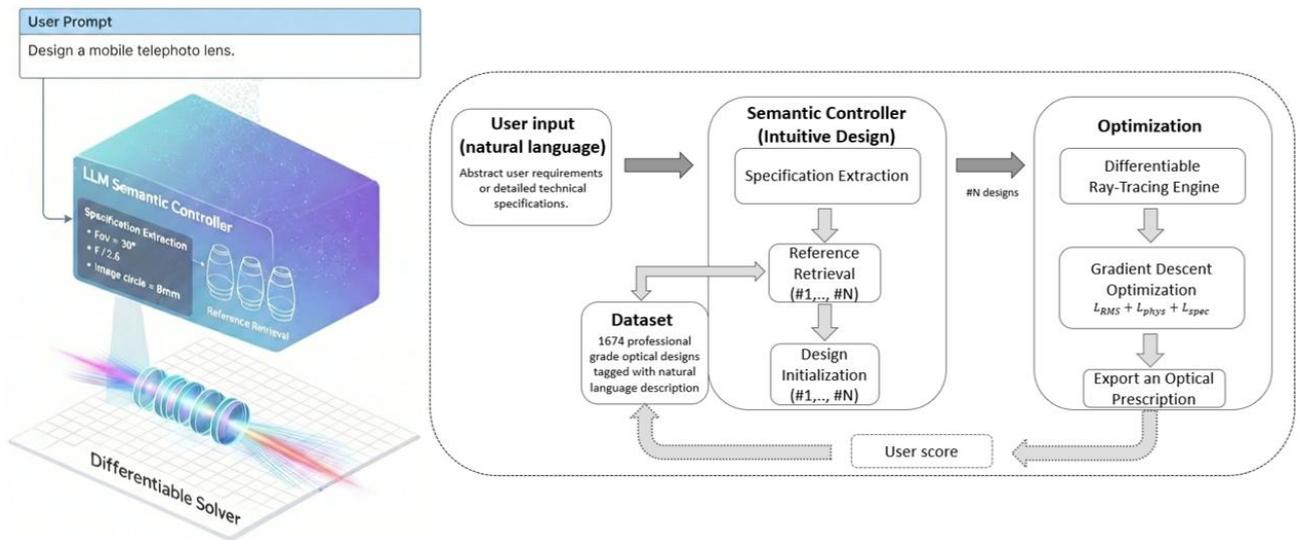

**Fig. 1 | Architecture of the physics-aware generative design pipeline.** *The workflow integrates semantic reasoning with differentiable physics to transform natural language into optical prescriptions.* **Left:** *Conceptual visualization of the system. An "LLM Semantic Controller" bridges the gap between abstract user prompts and physical constraints, generating "Intuitive Design Seeds" that serve as warm-start initializations for the physics engine.* **Right:** *The detailed algorithmic pipeline. The process begins with specification extraction, where the LLM interprets user constraints to query a reference library of 1,674 designs. The system retrieves the top N (typically N=3) closest matches and synthesizes a corresponding set of design candidates. These parallel initializations are then passed to a differentiable ray-tracing engine, which iteratively optimizes surface curvatures, glass thicknesses, and air spacings via gradient descent to minimize geometric aberrations and achieve the final target specification.*

The process begins when a user provides a high-level description (Fig. 1). Rather than trying to "draw" the lens immediately, the AI acts as a technical translator, mapping the user's intent to a primary vector of optical constraints. Specifically, the model extracts or infers four critical parameters: effective focal length (EFL), aperture size (F-number), sensor diagonal (image circle) and magnification (if applicable). Furthermore, the LLM identifies the specific application class (e.g., photographic prime) and the most appropriate optical archetype (e.g., telephoto). By converting subjective human intent into a structured, numerical query, the system enables a deterministic search of the design library, ensuring that the subsequent synthesis is grounded in a physically relevant starting point.

Once the primary engineering requirements are established, the system queries a curated reference library to ground the design in physical reality. We utilize the lens-designs.com database, a collaborative repository of approximately 1,700 validated optical prescriptions derived from patent literature and historical publications[41].

Using a deterministic search function, the pipeline identifies the three most functionally similar designs based on a weighted Euclidean distance of the extracted parameters. These designs serve as "expert demonstrations" within the LLM's context window. By analysing these reference prescriptions, the LLM performs analogical reasoning - interpolating between proven optical geometries to synthesize a novel starting prescription that satisfies the user's specific constraints. This strategy ensures the system avoids the "cold start" problem of traditional optimization, beginning instead from a physically grounded and high-performance design lineage. This part serves as an 'intuitive design' phase (Fig. 1), mimicking the heuristic-driven approach of a human designer before the system proceeds to numerical refinement.

For the optimization phase (Fig. 1), the synthesized designs are transferred to DiffOptics[1], an open-source differentiable ray-tracing library. Integrating a dedicated AutoDiff framework at this juncture allows the pipeline to move from a "sketch" to a diffraction-limited prescription by leveraging the power of gradient-based refinement.

The refinement process employs the Levenberg-Marquardt[42] (LM) algorithm within the DiffOptics environment to minimize a multi-objective loss function. As illustrated in Fig. 1, the total loss is formulated as a weighted sum:

$$L_{total} = \omega_{RMS} L_{RMS} + \omega_{phys} L_{phys} + \omega_{spec} L_{spec} \tag{1}$$

$L_{RMS}$ minimizes the root-mean-square (RMS) of the ray-intercept distributions on the image plane, driving the system toward sharp focus. $L_{phys}$ penalizes non-physical geometries, such as negative thicknesses, lens elements that overlap, or that are too thin to be physically manufactured. $L_{spec}$ ensures the final output adheres to the primary constraints extracted or inferred by the LLM, such as the target focal length, F-number, and sensor format.

Upon convergence, the synthesized prescriptions are exported to a standard format (ZMX) for independent verification. To ensure rigorous benchmarking, all optical layouts, RMS spot size diagrams, and Modulation Transfer Function (MTF) plots presented in this work are rendered using Ansys Zemax OpticStudio, validating the system's output against the industry-standard simulation environment.

**Generalization regimes: data interpolation and latent knowledge extrapolation**

The utility of the reference dataset within our pipeline is twofold. First, it serves as a repository of validated engineering archetypes, allowing the system to solve standard problems by retrieving and refining existing "best practices" rather than reinventing the wheel. Second, and perhaps more critically, the dataset acts as a "structural framing" device. While the LLM possesses extensive theoretical knowledge of optics from its pre-training, it often struggles to translate this abstract theory into a concrete numerical prescription. The dataset bridges this gap, providing the LLM with a practical schema (or "grammar") of optical surfaces, into which it can inject novel physical reasoning to satisfy constraints and ideas that do not exist in the training data.

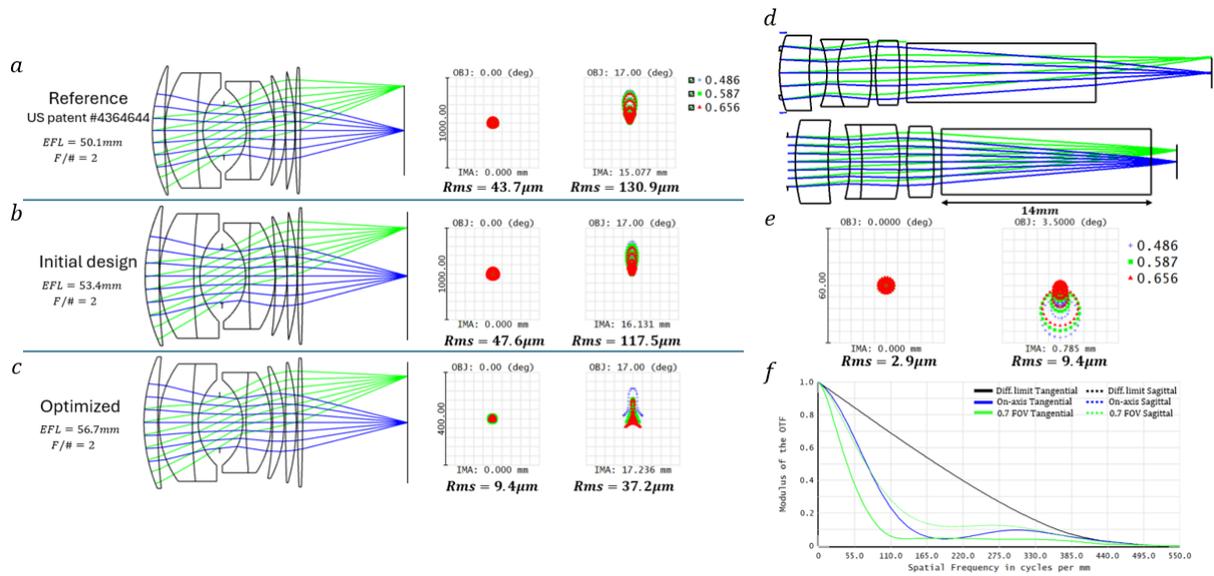

**Fig. 2 | Generative design validation: interpolation versus extrapolation. a–c, Interpolation within a dense design regime**. a, Optical layout and RMS spot size diagrams (on-axis and off-axis) for the retrieved reference design (US Patent #4364644), representative of a Double-Gauss photographic lens. b, Initial prescription synthesized by the LLM, correctly adopting the Double-Gauss topology. c, Final design after differentiable optimization, showing substantially reduced RMS spot sizes compared to both the LLM initialization and the original reference. **d–f, Extrapolation under novel geometric constraints.** d, Comparison of optical layouts for the LLM initialization (top) and the optimized retrofocus design (bottom), for a 10 mm f/3 lens with a mandatory 14 mm-thick glass prism inserted between the rear element and the image plane. This constraint enforces a back focal length exceeding the effective focal length, necessitating a retrofocus (inverted telephoto) topology. e, RMS spot size diagrams for the optimized system across the field. f, Modulation Transfer Function (MTF) performance, showing sagittal (solid) and tangential (dashed) contrast for on-axis and off-axis (70% FOV) field points. The system successfully synthesizes a structurally consistent retrofocus design despite the absence of explicit retrofocus examples or prism-coupled lenses in the reference dataset.

**Interpolation of standard archetypes.** To validate the system's ability to map standard requests to proven designs, we prompted the pipeline for a classic specification: *"Design a 50mm photographic lens at f/2."* This request falls squarely within the dense region of our parameter space (see supplementary note 1, figure 2).

The system correctly identified the request as a Double-Gauss derivative and retrieved a matching reference (fig.2a). Utilizing this reference as an initialization, the pipeline applied its differentiable optimization stage. The final design achieved a significantly lower RMS spot size than both the initial LLM output and the original reference patent (fig.2c). Consequently, the system proves capable of leveraging the dataset to bypass the 'cold start' problem, validating an end-to-end process that autonomously elevates retrieved historical designs to superior performance levels.

**Latent knowledge and novel constraints extrapolation.** The second experiment tested the system's capacity for independent synthesis when faced with specifications that lie outside the scope of the reference library. We prompted the pipeline with a geometrically unique request: *"Design a lens with a 10mm focal length, a 10° field of view, and an f/3 aperture, including a 14mm-thick glass prism inserted between the rear element and the image plane."*

This prompt imposes a constraint that necessitates a structural departure from standard configurations. While the requested focal length is short (10mm), the mandatory insertion of a 14mm physical prism forces the Back Focal Length (BFL) to significantly exceed the Effective Focal Length (EFL). This geometric condition inherently requires a retrofocus (inverted telephoto) architecture—a specific topology where a negative front group is paired with a positive rear group to shift the principal planes[43]. Crucially, the prompt did not contain the term "retrofocus," nor did the dataset contain examples of simple lenses coupled with a 14mm thick prism.

The results (fig.2d-f) demonstrate that the LLM successfully bridged the gap between retrieval and reasoning. It inferred a Triplet as the design basis but autonomously adapted the spacing and power distribution to accommodate the prism. By correctly placing the 14mm glass block in the optical train and adjusting the preceding elements, the system synthesized a design with clear retrofocus characteristics. Although the final optimized prescription drifted to an EFL of 13mm (a trade-off to maintain realizable curvature), the topology proved that the model could use the dataset to "frame" the optical surfaces while employing latent physical knowledge to satisfy a novel, constructive constraint.

**Synthesis of Novel and Practical Optical Systems**

The first case study (Fig. 3a) evaluates the system's ability to interpret a non-expert request for macro-scale electronic inspection, initialized by the prompt: *'I need a lens setup that lets me take really clear close-up pictures of electronic parts of 0.1mmx0.2mm. The camera is about 10 cm away from the object and I don't want the image to be distorted. I'm using a standard color camera.'*. The pipeline successfully identifies the "close-up" requirement as a finite-conjugate task, autonomously selecting a Double Gauss archetype - a symmetric topology inherently suited for minimizing distortion and field curvature at moderate magnifications. By inferring a common 1/2.3-inch sensor format (8.8 mm image circle), the LLM sets a target magnification of approximately $0.5 \times$ to ensure that the specified sub-millimeter features occupy a significant portion of the sensor area without requiring extreme, aberration-prone working distances.

The resulting synthesized prescription (EFL = 51.2 mm) converges to a working F-number ($F_w$) of 5.88 and a final magnification of $0.57 \times$. This configuration represents a physics-aware compromise: while a lower F-number would increase light throughput, the system's selection of $F_w \approx 6$ provides the necessary depth of field (DOF) to keep the 3D topography of electronic components in focus and ease the inspection (At this aperture, the design provides a total DOF of approximately $0.36 mm$)

Performance validation in Zemax (Fig. 3a, middle) reveals an on-axis RMS spot radius of $10.1 \mu m$. When compared to the theoretical Airy disk radius of $\sim 4 \mu m$ (at $\lambda = 550 nm$) the design falls short of the diffraction-limited performance at the field center for a first-pass synthesis. At 70% field position, the spot radius is increased to $22.1 \mu m$.

While falling short of a diffraction limited system, the design still provides a valuable solution to the described challenge. From a practical implementation standpoint, the utility of this design is best illustrated by its sampling density on a standard industrial sensor (e.g., $3.45\mu m$ pixel pitch). At a magnification of $0.57 \times$, a $0.1mm \times 0.2mm$ electronic component is projected onto the sensor as a $57\mu m \times 114\mu m$ image. This results in an effective resolution of approximately $16 \times 33$ pixels per component. Because the on-axis spot only spans roughly three pixels, the system avoids significant "blurring" across adjacent pixels, preserving the sharp edges required for automated optical inspection (AOI) of solder joints or component alignment.

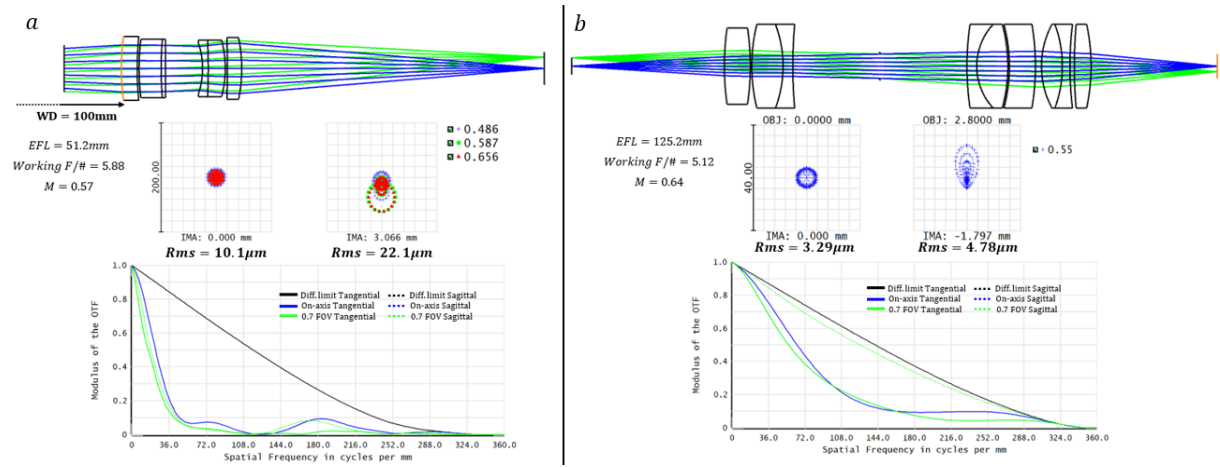

*Fig. 3 | Synthesis of finite-conjugate systems for industrial inspection. a, Macro-objective for high-resolution electronic component analysis synthesized from non-expert natural language. The layout (top) shows a symmetric Double Gauss configuration optimized for a 100 mm working distance (indicated by an axis break) and $0.57 \times$ magnification. RMS spot size diagrams (middle) demonstrate an on-axis radius of $10.1\mu m$, maintaining clarity across the field ($22.1\mu m$ at 70% FOV) for a working F-number of 5.88. The Modulation Transfer Function (MTF, bottom) confirms decent contrast for $0.1mm \times 0.2mm$ target features up to spatial frequencies of ~30 cycles/mm, providing sufficient sampling density for automated defect detection. b, Double-sided telecentric lens for precision mechanical metrology. The system adopts a specific topology (top) where the entrance and exit pupils are located at infinity, ensuring magnification remains invariant to object depth. Given a narrow-band illumination constraint, the design achieves near-diffraction-limited performance, with an on-axis RMS spot radius of $3.29\mu m$ matching the theoretical Airy disk radius of $3.44\mu m$ (middle). The MTF (bottom) exhibits near-ideal contrast across the field, enabling sub-pixel edge detection for accurate dimensional measurement of mechanical parts. See supplementary note 2 for initial and refence designs.*

The second case study (Fig. 3b) demonstrates the pipeline's capacity to synthesize a double-sided telecentric lens for high-precision industrial metrology, responding to the parametric request: *'Design a telecentric lens for inspecting mechanical parts from above. The object field is 8 mm, working distance around 50mm, magnification around 0.5x, and I want depth variation in the sample to minimally affect magnification. I'm using a narrow LED band.'*. The system correctly identifies that "minimizing magnification changes with depth" requires telecentricity in both object and image space - a structural requirement far more complex than standard imaging. The LLM reasoning determines that for an 8 mm object field at $0.5 \times$ magnification, a focal length of approximately 100 mm balances the requested 50 mm working distance with the necessary clearance for the internal aperture stop, which defines the telecentricity of the system.

The synthesized design (EFL = 125.2 mm) achieves a magnification of $0.64 \times$ and a working F-number of 5.12. By leveraging the user's constraint that chromatic aberration is non-critical due to narrow-band LED illumination, the optimization focuses entirely on achieving extreme field flatness and eliminating perspective error. The LLM assumed a wavelength of $550 nm$. The performance results are exceptional: the on-axis RMS spot radius of $3.29 \mu m$ is effectively at the theoretical diffraction limit (Airy radius is $\approx 3.44 \mu m$ at $\lambda = 550 nm$. Even at the 70% field position, the spot radius remains remarkably tight at $4.78 \mu m$.

The practical implications for mechanical part inspection are significant. Because the system is double-sided telecentric, the image of a part remains the same size even if the part is slightly displaced along the optical axis (depth variation). This orthographic projection is essential for metrology tasks, such as measuring the diameter of a screw or the spacing of mechanical slots, where perspective distortion in a standard lens would introduce measurement errors. The near-diffraction-limited MTF across the field ensures that the edges of mechanical parts are captured with high contrast, enabling sub-pixel edge detection algorithms to operate with maximum reliability in an automated factory environment.

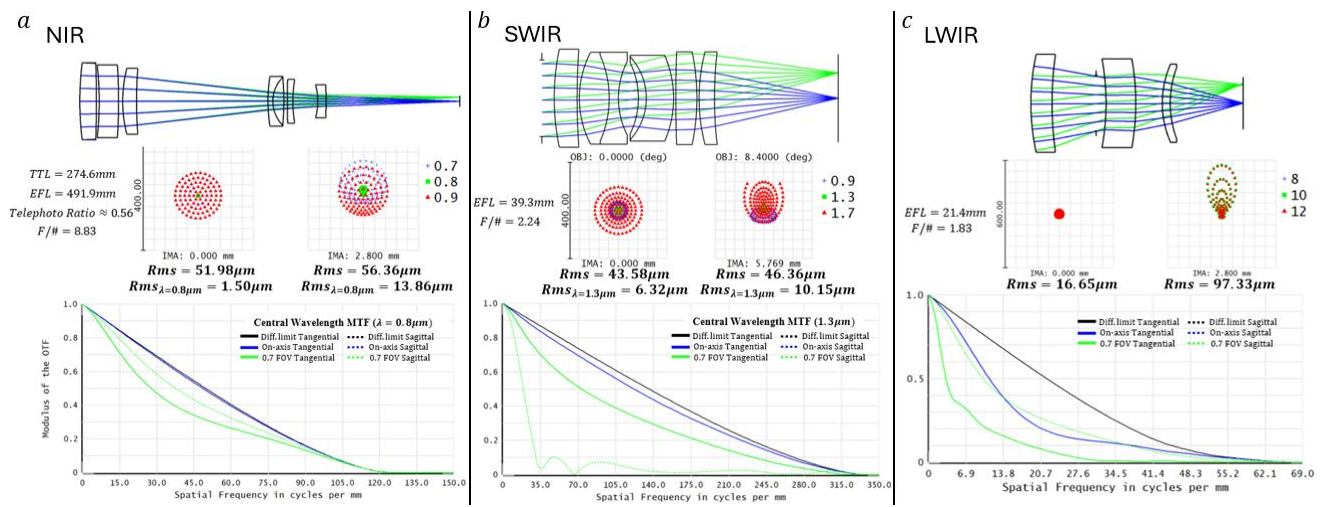

**Fig. 4 | Synthesis of infrared (IR) objectives and structural inference. a**, NIR telephoto objective (F/8.83) with a 500 mm focal length. The layout (top) demonstrates a successful structural synthesis of a positive-negative power distribution to achieve a telephoto ratio of 0.56, meeting compact form-factor requirements. RMS spot diagrams (middle) highlight diffraction-limited performance at the design wavelength (0.8 μm, 1.5 μm radius), while the polychromatic expansion (700–900 nm) reflects current material-model constraints in the optimizer. **b**, SWIR wide-aperture objective (F/2.24) for InGaAs sensors. The on-axis RMS spot radius (6.32 μm) remains significantly smaller than the targeted 15 μm pixel pitch, ensuring high energy concentration for the central design wavelength (1.3 μm). **c**, LWIR thermal imaging lens synthesized using Germanium substrates. The design is fully diffraction-limited on-axis, with an RMS radius (16.65 μm) smaller than the theoretical Airy disk radius (22.32 μm at $\lambda = 10 \mu m$). Notably, the spot sizes across the 8–12 μm band remain consistent due to the inherent achromaticity of Germanium in the thermal regime, which mitigates the dispersion-modeling challenges observed in shorter IR bands. The average MTF (bottom) confirms the pipeline's ability to generate physics-ready prescriptions for specialized thermal monitoring. See supplementary note 2 for initial and refence designs.

To test the system's capacity for original structural synthesis, we prompted it to design "*a telephoto lens for the NIR spectral range (700-900nm). Focal length 500mm, F/10, telephoto ratio less than 0.6. Image format 1/2.*". For this extreme practical requirement, the model cannot rely on retrieved references, as the specified total track length (TTL < 300 mm) for such a long focal length requires a highly specific power distribution. The LLM correctly inferred that achieving a compact form factor necessitates a classic telephoto topology: a high-power positive front group followed by a divergent rear group to shift the principal plane forward.

The synthesized design (Fig. 4a) achieves a focal length of 491.9 mm within a physical length of only 274.6 mm, resulting in a telephoto ratio of 0.56. This exceeds the user's compactness requirement while maintaining an aperture of $F/8.83$. At the primary wavelength of $800nm$, the design exhibits exceptional performance with an on-axis RMS spot radius of $1.5\mu m$. When compared to the theoretical Airy disk radius of $8.6\mu m$, the system is comfortably diffraction-limited. Even at the 70% field position, the monochromatic RMS radius remains tight at $13.86\mu m$.

We observe, however, a significant increase in the poly-chromatic RMS spot size ($51.98\mu m$ on-axis across the 700–900 nm band). This is attributed to a current limitation in the differentiable optimization engine, which presently utilizes simplified refractive index and Abbe number models rather than full wavelength-dependent dispersion descriptions based on Sellmeier coefficients for out-of-visible-range materials. Consequently, while the pipeline successfully synthesized a high-performance monochromatic topology, the correction of the primary axial chromatic aberration and lateral chromatic aberration remains a point for future refinement. Nevertheless, the monochromatic MTF at the central wavelength (Fig. 4a, bottom) remains near ideal.

Expanding the pipeline's spectral reach into the Short-Wave Infrared (SWIR, 0.9–1.7 µm) regime, we synthesized a wide-aperture objective in response to the explicit request: *'I need a wide-aperture short-wave IR imaging objective for a 640x512 InGaAs sensor. Target focal length ~35 mm, f-number <=2, and FOV of at least $\pm$12°. Prefer 6-8 elements.'*. The system correctly identified that the requested wide field of view (12° half-field) necessitated a modified Double Gauss architecture. The synthesized objective (EFL = 39.3 mm, F/2.24) demonstrates the model's ability to navigate the complex trade-offs between aperture size and spatial resolution. At the central design wavelength of $1.3\mu m$, the system achieves an on-axis RMS spot radius of $6.32\mu m$, while the theoretical Airy disk radius is 3.56µm This high level of correction persists across the field, with an RMS radius of $10.15\mu m$ at the 70% field position. As with the NIR telephoto case, the observed polychromatic RMS expansion ($43.58\mu m$) stems from the current optimizer's reliance on simplified $n/v$ material modeling rather than comprehensive SWIR dispersion coefficients - a limitation that primarily affects the secondary spectrum while leaving the underlying monochromatic topology highly optimized.

The final case study in our infrared suite (Fig. 4c) addresses the Long-Wave Infrared (LWIR, $8-12\mu m$) spectrum, responding to the specification: *'I would like to design a thermal imaging lens for the LWIR spectral range (8-12 microns). Focal length 25mm, F/2, image diagonal 8mm.'*. The LLM selected Germanium (Ge) as the primary substrate. successfully inferring its high refractive index and low dispersion as the ideal solution for a compact, high-speed thermal objective.

The synthesized prescription (EFL = 21.4 mm, F/1.83) demonstrates the robust monochromatic performance achievable in the thermal band. Unlike the previous NIR and SWIR examples, the LWIR design shows a remarkable consistency across its three design wavelengths (8, 10, and $12\mu m$). This is due to the physical properties of Germanium: in the LWIR band, Germanium has an exceptionally high Abbe number, meaning its refractive index changes very little with wavelength. Consequently, the dispersion-modeling limitations of the current optimizer, which challenged the polychromatic performance in shorter IR bands, are physically mitigated here by the material's inherent achromaticity.

The performance validation (Fig. 4c, middle) reveals an on-axis RMS spot radius of $16.77\mu m$. When compared to the theoretical Airy disk radius of $\sim 22\mu m$, the design is fully diffraction-limited at the center of the field. The off-axis spot radius increases to $95.93\mu m$ at the 70% field position. The average MTF (Fig. 4c, bottom) remains robust for the lower spatial frequencies.

**Challenges in High-Order Aspheric Synthesis**

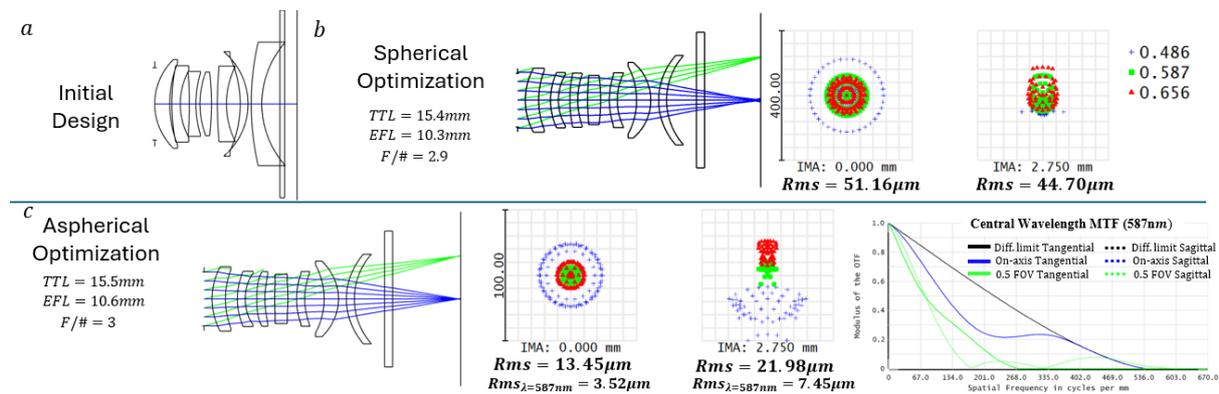

*Fig. 5 | Optimization challenges and staged solution for aspheric mobile lenses. a, Raw initialization of a 1G6P (1 Glass, 6 Plastic) hybrid lens generated by the LLM. The output identifies the correct topology but exhibits overlaps and ray failures. b, Result of Phase 1 (Geometric Stabilization). By restricting the solver to spherical terms, the system converges to a valid, "healthy" ray trace state. This resolves physical conflicts but results in a relaxed form factor (TTL 15.4 mm, F/2.9) with high residual aberrations (RMS ≈ 50μm. c, Result of Phase 2 (Aspheric Refinement). Using a staged release of coefficients (conic → $\alpha_4$ … → $\alpha_8$) to handle the non-linear optimization landscape, the system corrects the wavefront while maintaining the stable geometry from Phase 1. This reduces on-axis monochromatic RMS to 3.52μm (MTF 0.43 at 125 lp/mm).*

To probe the upper boundaries of the pipeline's capabilities, we tasked the system with designing a *"flagship mobile wide-angle lens"* (Fig. 5). The prompt imposed the rigorous constraints typical of modern smartphone optics: *"a 1G6P hybrid configuration (one glass, six plastic elements) for a 200MP sensor, an effective focal length of 6.9mm with a fast aperture (F/1.7) and an ultra-compact Total Track Length (TTL) of 7.5 mm."*

The LLM successfully retrieved the correct architectural prior, proposing a valid 8-element sequence (1 glass + 6 polymer aspheres + cover glass). However, as is common with zero-shot synthesis of highly dense systems, the raw output contained element overlaps and ray failures (Fig. 5a). These violations are explicitly penalized by the physical regularization term $L_{\text{phys}}$ in Eq. (1), which drives the optimization toward a physically realizable geometry. Dealing with these violations introduces a trade-off: the optimizer prioritizes physical feasibility over strict adherence to the prescribed total system length, leading to an increase in overall optical length.

The optimization of high-order polynomial surfaces is notoriously non-linear. Attempting to solve for geometry and aspheric coefficients simultaneously often drives the solver into local minimum or generates invalid, imaginary sag profiles. To resolve this, we employed a multi-phase curriculum optimization strategy designed specifically to handle the ill-posed nature of aspheric lens design.

**Phase 1 (Geometric Stabilization):** We first restricted the optimization to spherical curvatures and element spacings only. This step prioritizes physical realizability over wavefront correction, allowing the system to converge into a "healthy" geometric state where rays successfully trace through the aperture to the sensor. The output of this phase is shown in Fig.5b. The solver favored a relaxed mechanical state to minimize ray error, resulting in a Total Track Length (TTL) of 15.4 mm and an F-number of 2.9. Although the aberrations remained high ($RMS \approx 50 \mu m$), this phase established the necessary convex basin for subsequent refinement.

**Phase 2 (Aspheric Refinement):** With the ray paths stabilized, we locked the global geometry (curvatures and thickness) and progressively released the aspheric degrees of freedom. Because aspheric coefficients ($k, \alpha_4, \alpha_6, \alpha_8$) operate at vastly different numerical magnitudes, we adopted a staged release protocol. We optimized the conic constants first, followed by $4^{th}$ order terms, and finally up to $8^{th}$ order coefficients, allowing the solver to fine-tune previous terms at each step. This strategy yielded a dramatic recovery in optical performance (Fig. 5c). The polychromatic RMS spot size dropped to $13.45 \ \mu m$, and for the primary design wavelength (587 nm), the on-axis RMS spot radius reached $3.52 \ \mu m$. While the final design retains the relaxed mechanical dimensions and exhibits vignetting beyond the 0.5 field, the result confirms that this two-phase approach successfully solves the ill-posed problem of converting a semantic blueprint into a high-performance aspheric prescription.

Performance analysis of the final design indicates that chromatic aberration remains the dominant residual error, significantly limiting the MTF at the field edges. This spectral limitation is a direct consequence of the model's autonomous material selection for the 1G6P stack: N-LAK34 ($n_d = 1.729, V_d = 54.5$) for the front glass element, paired with OKP4HT ($V_d = 23.3$) and E48R ($V_d = 56.0$) for the subsequent plastic layers. While the high refractive index of N-LAK34 successfully reduced the total track length, its Abbe number is nearly identical to that of the crown plastic (E48R). This lack of a significant dispersion differential ($V_{glass} \approx V_{plastic}$) prevented the glass element from functioning as an effective achromatizing anchor, effectively leaving the highly dispersive OKP4HT elements without a strong counter-corrective flint partner in the glass layer.

## Discussion

We have presented a generative framework that bridges the semantic gap between natural language intent and rigorous optical engineering. By coupling the reasoning capabilities of Large Language Models with a differentiable ray-tracing engine, the system successfully synthesized valid starting points for diverse optical regimes—from finite-conjugate macro lenses to telecentric metrology systems and infrared objectives. This hybrid approach solves the "blank page problem," effectively automating the most intuitive and experience-dependent phase of optical design: the selection of a topological starting point.

**Data Scalability and Generalization**

The performance of the generative module is currently bounded by the scale and diversity of the training data. Our dataset, while sufficient to demonstrate transfer learning across standard imaging domains, remains limited in both volume and novelty compared to the vast repositories of proprietary industrial designs. As with broader trends in generative AI, we anticipate that scaling the dataset to include broader catalogs of patents and legacy designs will significantly enhance the model's ability to retrieve increasingly complex architectures. A larger, more diverse training corpus would likely mitigate the "geometric hallucinations" observed in the mobile lens initialization (Fig. 5a) by providing higher-fidelity structural priors.

**Physical and Optimization Constraints**

While the differentiable physics engine (*DiffOptics*) proved robust for rotationally symmetric refractive systems, several architectural limitations remain. Currently, the engine does not support reflective elements (mirrors) or coordinate breaks (tilted/decentered components), precluding the system's applicability to catadioptric telescopes or folded-path architectures. Furthermore, the optimization challenges observed in the IR and Mobile Lens cases highlight the need for higher-fidelity material modelling and stricter boundary constraint mechanisms. The polychromatic spot expansion in the NIR/SWIR designs (Fig. 4) stems from simplified refractive index modelling, suggesting that integrating full dispersion curves into the differentiable graph is a critical next step. Similarly, the mechanical relaxation of the mobile lens design (Fig. 5b) indicates that future iterations must incorporate rigid mechanical CAD constraints, such as fixed Total Track Lengths (TTL), directly into the loss function to prevent the solver from "trading" physical compactness for ray convergence.

**Future Horizons: Diffractive/Meta Optics and Afocal Systems**

The modular nature of the differentiable pipeline opens significant avenues for expansion. For instance, the current loss functions are tailored for focal imaging systems (minimizing spot size on a sensor); extending the framework to afocal systems (e.g., beam expanders, rifle scopes) would require the formulation of new angular-domain loss functions. This highlights the democratization potential of the approach: because the physics engine is differentiable, end-users can define custom merit functions for specialized tasks without redesigning the core optimizer. Furthermore, the framework is theoretically compatible with general phase objects, allowing for future integration of diffractive optical elements (DOEs) and metasurfaces. This would enable the synthesis of hybrid refractive-diffractive systems, further miniaturizing the designs for portable electronics and AR/VR applications.

**Conclusion**

This work demonstrates the first end-to-end pipeline for translating natural language directly into optimized optical prescriptions. By moving beyond simple parameter retrieval and enabling the physics-aware refinement of generative outputs, we have shown that AI can act not merely as a catalog but as an active participant in the engineering loop. While current limitations in dataset size and simulation fidelity define the present boundary of the technology, the successful synthesis of diffraction-limited designs across multiple spectral bands validates the core thesis: that the fusion of semantic reasoning and differentiable physics is the path toward autonomous optical engineering.

## Methods

### Dataset and Preprocessing

The generative model was trained on a comprehensive dataset of optical designs curated by Reiley and Claff[41]. The raw data, originally in ZMX format, was parsed to extract key optical parameters (surface curvatures, thicknesses, material definitions, and semi-diameters) and converted into a structured JSON format compatible with the Large Language Model (LLM) tokenization scheme. No proprietary or confidential industrial designs were utilized in the training process.

### Differentiable Optimization Framework

The physics-based refinement module was built upon *DiffOptics*[1], a differentiable ray-tracing engine enabling gradient-based optimization of optical systems. The framework was implemented in Python using the PyTorch library, allowing for the automatic differentiation of ray paths with respect to surface parameters (curvature, thickness, and aspheric coefficients). Optimization was performed using the LM algorithm[42] with a curriculum-based learning rate schedule, stabilizing the convergence of high-dimensional variables (such as the aspheric orders) as described in the mobile lens case study. Material dispersion was modelled using refractive index and abbe number (see discussion).

### Validation and Benchmarking

All synthesized optical designs were exported to Ansys Zemax OpticStudio (2023 R1 Premium)[44] for independent validation and performance characterization. This step ensured that the results were not artifacts of the differentiable approximation but physically realizable prescriptions. The performance matrix (EFL, TTL, F/#, spot size RMS, FFT MTF) were calculated using Zemax. To ensure proper axial placement of the image plane, the built-in *Quick Focus* optimization was applied once prior to evaluation. No additional Zemax optimization tools were used. Reference Airy disk radii were calculated within Zemax based on the working F-number ($F_w$) and the primary design wavelength. Simulations utilized standard industry glass catalogs to ensure manufacturability.